\documentclass[11pt,preprint]{aastex}

\newcommand{\HII}{\mbox{H\,{\sc ii}}}
\newcommand{\HI}{\mbox{H\,{\sc i}}}

\newcommand{\twCO}{$^{12}$CO}

\newcommand{\kmps}{\mbox{${\rm km\;s^{-1}}$}}

\newcommand{\lsim}{\mbox{$\mathrel{\vcenter{\hbox{\ooalign{\raise3pt\hbox{$<$}\crcr \lower3pt\hbox{$\sim$}}}}}$}}
\newcommand{\gsim}{\mbox{$\mathrel{\vcenter{\hbox{\ooalign{\raise3pt\hbox{$>$}\crcr \lower3pt\hbox{$\sim$}}}}}$}}
\shorttitle{Faint OH Emission from the Local ISM}
\shortauthors{Ronald J.\ Allen et.al.}

\begin{document}


\title{
Faint Extended OH Emission from the Local Interstellar Medium \\
in the Direction $l \approx 108^{\circ}, b \approx 5^{\circ}$ \\}


\author{Ronald J.\ Allen}
\affil{Space Telescope Science Institute, 3700 San Martin Drive, 
Baltimore, MD 21218}
\email{rjallen@stsci.edu}


\author{M\'onica Ivette Rodr\'{\i}guez}
\affil{Instituto de Astrof\'{\i}sica de Andaluc\'{\i}a, PO Box 3004, 18080 Granada, Spain}
\email{mrm@iaa.es}


\author{John H.\ Black}
\affil{Onsala Space Observatory, Onsala, SE-439 92 Sweden}
\email{John.Black@chalmers.se}


\author{Roy S.\ Booth}
\affil{Onsala Space Observatory; Onsala, SE-439 92 Sweden, and \\ SKA, Cape Town; SKA - SA, The Park, Park Road, Pinelands, 7405, South Africa}
\email{roy@hartrao.ac.za}



\begin{abstract}
We have mapped faint 1667 OH line emission ($T_A \approx 20 - 40$ mK in our $\approx 30'$ beam) along many lines of sight in the Galaxy covering an area of $\approx 4^{\circ} \times 4^{\circ}$ in the general direction of $l \approx 108^{\circ}, b \approx 5^{\circ}$. The OH emission is widespread, similar in extent to the local \HI\ ($r \lesssim 2$ kpc) both in space and in velocity. The OH profile amplitudes show a good general correlation with those of \HI\ in spectral channels of $\approx 1$ \kmps; this relation is described by $T_A$(OH) $ \approx 1.50 \times 10^{-4}\; T_B$(\HI) for values of $T_B$(\HI) $ \lesssim 60 - 70$ K. Beyond this the \HI\ line appears to ``saturate'', and few values are recorded above $\approx 90$ K. However, the OH brightness continues to rise, by a further factor $\approx 3$. The OH velocity profiles show multiple features with widths typically 2 -- 3 \kmps, but less than $10\%$ of these features are associated with CO(1-0) emission in existing surveys of the area smoothed to comparable resolution. 
\end{abstract}


\section{Introduction}
\label{sec:intro}

The initial detection of the 18-cm radio lines of OH was achieved more than 48 years ago \citep{wbmh63} in absorption against the continuum emission from the bright Galactic radio source Cas A, and some of the first information on the Galactic distribution of OH was provided by early surveys of absorption against Galactic continuum sources \cite[e.g.][]{gos68}. The anomalous OH maser emission was serendipitously discovered in targeted searches towards Galactic \HII\ regions in 1965 \citep{wwdl65, mrgb65}. This remarkable discovery rapidly transformed studies of OH in the ISM to focus more narrowly on specific sources in an attempt to understand the nature and physical origin of the anomalous excitation \cite[e.g.][]{th71, cru73, tu73, rwg76, ger78, dck81}. The subject has by now become quite mature with an extensive literature \cite[for reviews see e.g.][]{rm81, eli92}.

Early ``blind'' searches for extended OH emission from the general ISM in the Galaxy were unsuccessful \citep{pen64}, even at noise levels down to $\sigma_{rms} \approx 20-30$ mK \citep{kk72}, and the general conclusion was that ``normal'' OH emission\footnote{In this context, ``normal'' refers to emission with LTE line ratios of 5/9 in the main lines at 1665/1667 MHz. However, the excitation temperature of this emission is generally subthermal, i.e.\ below the kinetic temperature of the gas in which the OH molecules are embedded \cite[e.g.][]{gos68}.} from the extended ISM could be found only in regions of high dust or gas density \cite[e.g.][]{hei68}. However, in their study of physical conditions in OH-emitting/absorbing clouds, \citet{dck81} pointed out that faint OH emission may nevertheless be widespread in the diffuse ISM. The excitation temperature of this gas in the main lines was by that time known to be high enough \cite[$\gtrsim 6$ K, e.g.][]{rwg76} that it would be largely absent in OH absorption spectra, appearing only weakly against moderately bright background sources. The way to reveal the possible presence of such a ubiquitous component would be to conduct a deep emission survey along lines of sight away from any background sources. Such a survey has not yet been done, and the general large-scale Galactic distribution of OH emission is still unknown.

As receiver sensitivity continued to improve throughout the 1980's, faint OH emission with a wide velocity extent was occasionally reported in the community of OH observers. This phenomenon appeared as perturbations in the baselines on high-sensitivity spectra of specific Galactic objects, and such features were often  inadvertently removed from modest-sensitivity observations by the baseline subtraction process. A good example of this is the OH Zeeman study of the molecular cloud Barnard 1 (B1) by \citet{goo89} using the Arecibo radio telescope. The 1665 and 1667 OH emission profiles of B1 in that study are shown in their Figure 1; both profiles peak at $\approx +6$ km/s, but both also show a faint extension at a level of $\approx 30$ mK (in Stokes I) towards lower velocities of $\approx +2$ km/s. The noise level in these data is extremely low ($\approx 5$ mK), a consequence of system temperature improvements and the long integration times (30-40 hrs) required on these targets in order to detect the Zeeman signature of OH. There is therefore little doubt that these extended features are real, although the paper by \citet{goo89} does not discuss them\footnote{We are grateful to A.A.\ Goodman for drawing our attention to this result.}.

The first quantitative results on faint extended OH emission features in the Galaxy were reported by \citet{lilu96} in their study of OH emission and absorption along lines of sight towards a sample of 17 compact extragalactic continuum sources. Their high-sensitivity, wide-band ``expected'' OH emission profiles, constructed from many spectra taken near (but not on) the background sources, is a ``miniature'' version of the survey suggested by \citet{dck81}. \citet{lilu96} recorded faint OH emission profiles with a wide velocity extent in more than half of the directions studied (cf.\ their Figure 1). One especially dramatic case is their OH spectrum of the source 0355+508, which also appears (along with other molecular tracers) in Figure 6 of a follow-up paper by these same authors \citep{lilu00}.

Here we present the results from another ``miniature'' OH emission survey which we have conducted, somewhat serendipitously, as part of our study of radio line and continuum emission from the large Galactic dust cloud Lynds 1204 \cite[][]{rod07}. Initially focused narrowly on detecting faint OH emission associated with the dust cloud itself (which, according to \citet{lyn62}, covers an area of $\approx 2.5$ square degrees), our survey was progressively enlarged to cover an area of $\approx 16$ square degrees in a vain attempt to find emission-free regions around L1204 which could serve as ``blank sky''. In the end, the profiles recorded at the location of L1204 are entirely similar to those found elsewhere in the survey area well away from it. Our 1667 MHz OH emission ``mini-survey'' is described in \S \ref{sec:observations}. In \S \ref{sec:results} we compare the major morphological features of our OH survey with existing \HI\ and CO surveys of the same area. In \S \ref{sec:conclusions} we summarize our conclusions and outline several suggestions for future work.

\section{Observations}
\label{sec:observations}

\subsection{Equipment and data acquisition}

The observing program was carried out over a period of several months in late 2005 with the 25m radio telescope of the Onsala Space Observatory. The main characteristics of this radio telescope at 18cm wavelength have hardly changed since the observations by \citet[][]{gwj76} and \citet[][]{rwg76}. They are: main beam FWHM $27' \times 31'$ ($\alpha \times \delta$); main beam efficiency $64 \pm 8\%$, aperture efficiency $52 \pm 2\%$. Both polarization states were observed and the data averaged. The receiver front end system temperature is $T_{sys} \approx 25 - 35$ K. A digital correlation spectrometer is now in use with 800 channels spread over a bandwidth of 3.2 MHz, so the spectral resolution is 4.0 kHz = 0.72 \kmps\ at 1667.3590 MHz\footnote{This was subsequently smoothed to $\approx 1$ \kmps\ during an interpolation intended to bring the OH profiles in velocity alignment with the corresponding HI profiles.}. The receiver central frequency was switched, and the coverage was such that both the 1665 and the 1667 OH main lines were initially measured in the difference spectra. Unfortunately the 1665 data were often plagued by interference, and the latter parts of the survey focused only on the 1667 line. The final spectra are given in units of $T_A$.

\subsection{Survey construction}

\begin{figure*}[ht!] 
\includegraphics[width=\textwidth, angle=0]{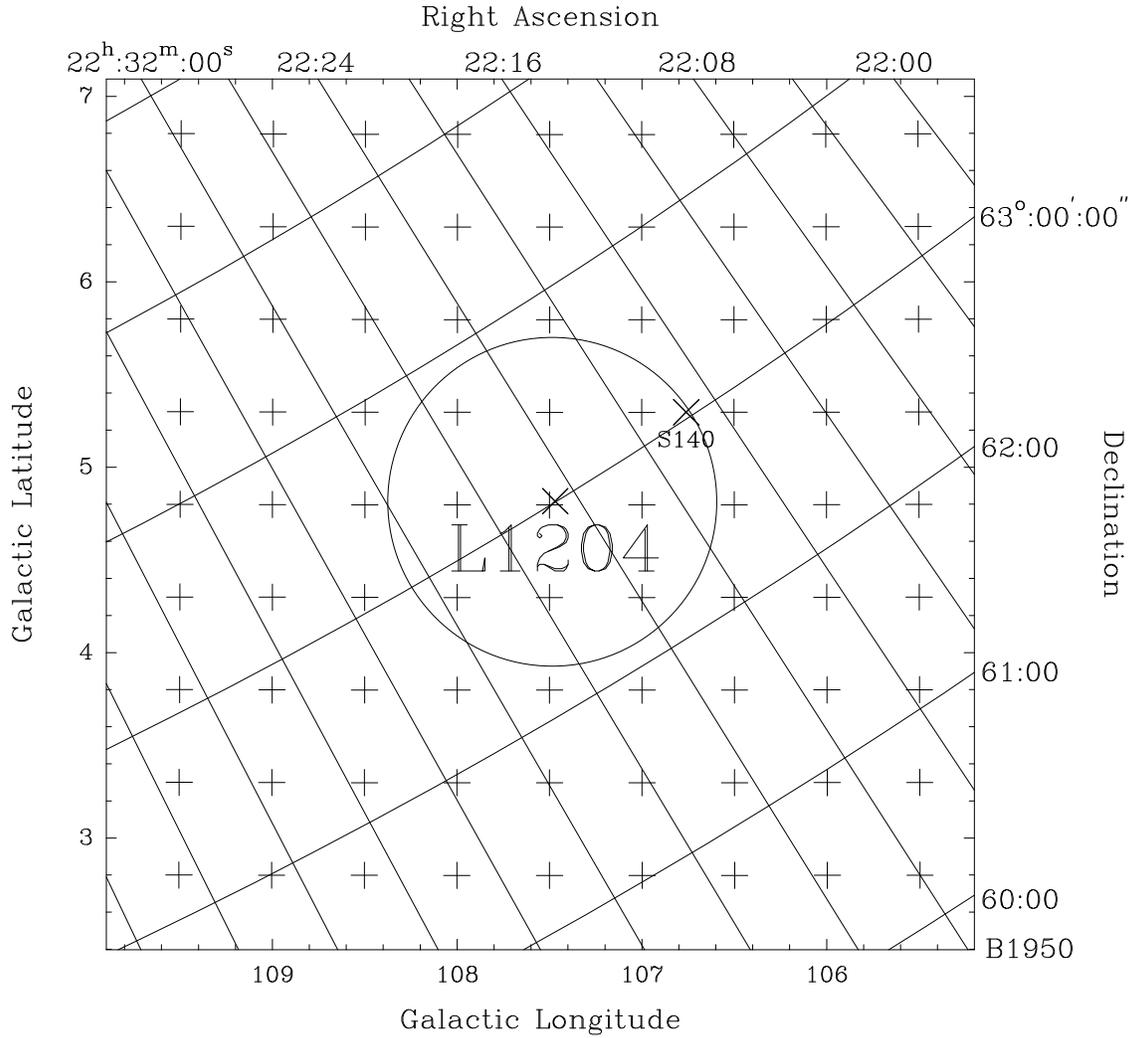}
\caption{
Sketch of the survey area. The grid of profiles is oriented in Galactic longitude and latitude, centered at $l = 107.5^{\circ}, b = +4.8^{\circ}$. The grid separation is $30'$, which is closely equal to the telescope beam FWHM. The approximate location of the dust cloud Lynds 1204 \cite[$l = 107.47^{\circ}, b = +4.82^{\circ}$;][]{lyn62} is also shown. The circle drawn around this location is meant to show the approximate area of the cloud (2.5 square degrees) according to Lynds, not the actual boundary (which is very irregular). Also shown is the location of the H$\alpha$ bright-rim nebula S140 \cite[$l = 106.8^{\circ}, b = +5.3^{\circ}$;][]{s59}. Note that the equatorial coordinates in this figure are for epoch B1950.
}
\label{fig:survey}
\end{figure*}

The initial survey in October 2005 was aimed at observing OH emission in Lynds 1204 and, since the dust cloud was thought to be less than $\approx 1.8^{\circ}$ in diameter, the observing was done on a regular $5 \times 5$ grid of points oriented in Galactic longitude and latitude, centered at $l = 107.5^{\circ}, b = 4.8^{\circ}$, and separated by $30'$. Total integration times per point were typically a few hours, made up of many shorter observations. We (unexpectedly) continued to record detectable OH profiles at the outer boundaries of the grid, so in the December observations we enlarged the grid to a $9 \times 9$, well beyond the extent of the dust cloud L1204 as recorded by Lynds, and covering a total area of $\approx 4^{\circ} \times 4^{\circ}$. A sketch of the observing grid and of the approximate boundary of L1204 is shown in Figure \ref{fig:survey}.

%

\subsection{Data reduction}

The OH profiles recorded during our survey are faint, and the radio spectrum at these wavelengths is often plagued with interference generated both external and internal to the observatory. Our observing run was no exception to this unfortunate rule. Interference from an external, frequency-unstable source occasionally affected the receiver, at strengths sufficient to perturb the baselines in our frequency-switched spectra. Each profile was examined visually and removed from the data set if these perturbations were considered unacceptable.  Approximately 15\% of the data acquired were discarded for this reason. In addition, faint but frequency-stable interference was also present at various discrete frequencies within the correlator passband. The spectrum around the 1665 line was sufficiently disturbed by this interference that further reduction of that data was suspended. The repeatability of the 1667 MHz profiles was checked by returning to the central (0,0) reference point of the observing grid at regular intervals throughout the observations and recording a profile with the same integration time as that adopted for all grid points. The spectra recorded at this reference point were later collected and analyzed for internal consistency and the effects of low-level interference; the results of that analysis are give in Appendix \ref{app:one}. In addition, the OH emission profile from the strong maser source W3OH was measured at the beginning and at the end of each observing session as an overall check on system operation.

Each remaining 1667 MHz frequency-switched OH profile was ``folded'' and averaged, visually examined, and further processed using the XS profile analysis system developed at Onsala. Linear baselines were fitted over the velocity ranges $-45 \leq V_{LSR} \leq -25$ \kmps\ and $25 \leq V_{LSR} \leq 45$ \kmps\ (see Appendix \ref{app:one} for the justification of this choice) and subtracted, and all profiles at each survey position were then averaged together.

\section{Survey results}
\label{sec:results}

\begin{figure*}[ht!] 
\includegraphics[width=6.0in, angle=-90]{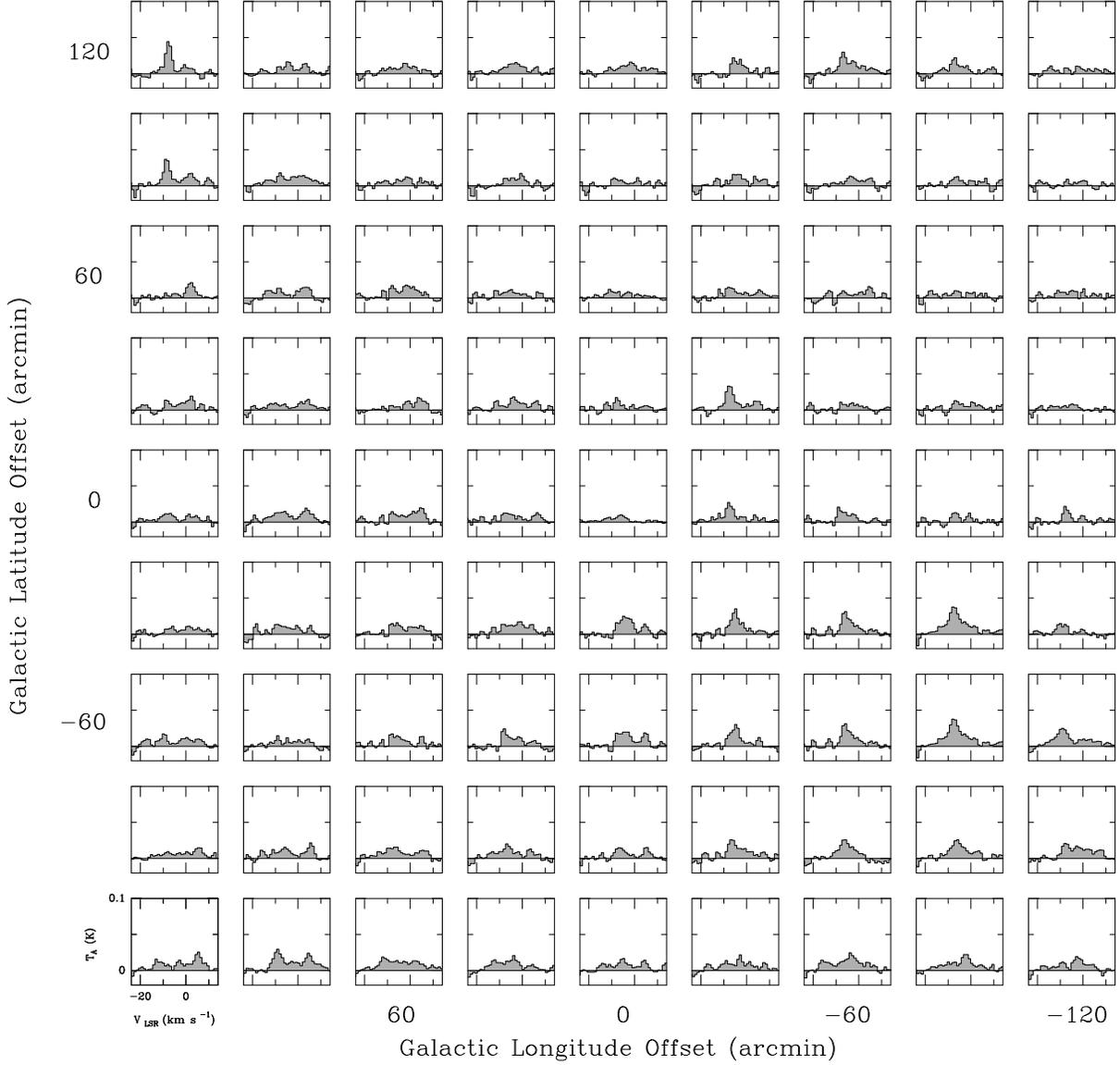}
\caption{
Mosaic of OH profiles recorded by our survey. Antenna temperature ($T_A$) and LSR velocity scales are given on the panel at the lower left corner. The grid of profiles is oriented in Galactic longitude and latitude, and centered at $l = 107.5^{\circ}, b = 4.8^{\circ}$. The grid separation is $30'$, which is closely equal to the telescope beam FWHM. The velocity resolution in this plot has been smoothed to $\approx 1$ \kmps\ for comparison with the \HI\ profiles in Figure \ref{fig:HI}.
}
\label{fig:OH}
\end{figure*}

The 1667 MHz OH profiles in our survey area are displayed in the mosaic of Figure \ref{fig:OH}. The total observing time for each position shown here is typically 3 hours after removal of scans deemed to be of insufficient quality. The signal peaks are generally about $T_A \approx 20 - 40$ mK, and the noise levels in these final profiles as established from the baseline fits vary from 3 to 5 mK in $T_A$ units. The central pixel at offset (0,0) was observed more often (total 10.0 hours) as a check on system performance; the noise level on this profile is 2.0 mK (see Appendix \ref{app:one} for further discussion of this profile).

One of the most striking results of our survey is the ubiquity of the OH emission. As can be seen from Figure \ref{fig:OH}, significant OH profiles were recorded at all survey positions, and roughly with the same levels of intensity. This may at first sight seem surprising if one is used to thinking of molecular clouds in the ISM as more isolated concentrations of gas primarily associated with regions of recent star-formation activity. The wide spatial extent of the OH emission is matched by the broad extent of the profiles in velocity. The profiles also show significant ``substructure'', each profile consisting of several peaks each with a signal/noise ratio in excess of 4. We discuss each of these points further in this section. 

\subsection{Correspondence of OH with \HI}

\begin{figure*}[ht!] 
\includegraphics[width=6.0in, angle=-90]{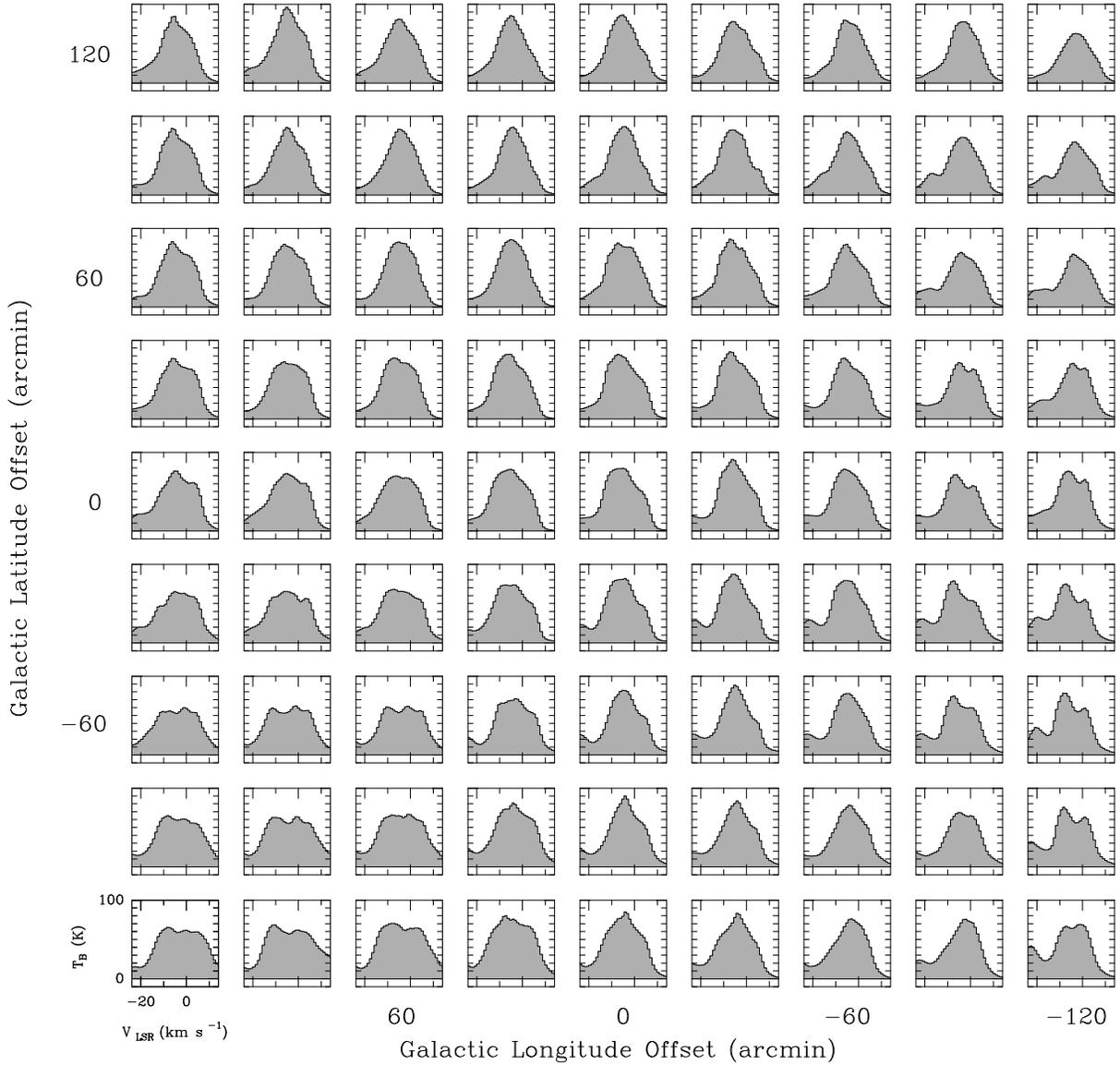}
\caption{
21-cm \HI\ profiles in the survey area. Brightness temperature ($T_B$) and LSR velocity scales are given on the panel at the lower left corner.
}
\label{fig:HI}
\end{figure*}

The spatial and velocity extents of the OH profiles in our survey resemble those of the \HI. In order to illustrate this point we have constructed \HI\ profiles at each position of our survey using data from the ``Atlas of Galactic Neutral Hydrogen'' \citep{ha97}. This Atlas collects the spectra of the Leiden/Dwingeloo HI survey that was carried out using a 25-m telescope. The angular resolution of that telescope at 21 cm is also $\approx 30'$, matching closely with our OH observations, so no additional spatial smoothing was applied. The full velocity range of the \HI\ survey covers from -450 to +400 \kmps, the spectral resolution is about 1.0 \kmps, and the brightness temperature sensitivity is $\approx 0.07$ K. The observations in the Atlas are presented as (latitude, velocity) maps at longitude intervals in steps of $30'$; in order to compare with our survey data, we have extracted the \HI\ profiles from the Atlas at the appropriate positions and velocity interval and plotted them on our survey grid using the GREG package from the GILDAS software system \citep{gf89}. The resulting mosaic of \HI\ profiles is shown in Figure \ref{fig:HI}. As we have asserted earlier, the spatial and velocity extent of the 1667 MHz OH emission is very similar to that of the 21-cm \HI\ emission.

\subsubsection{A detailed comparison of OH and HI profiles}

\begin{figure*}[ht!] 
\begin{center}
\includegraphics[width=\textwidth]{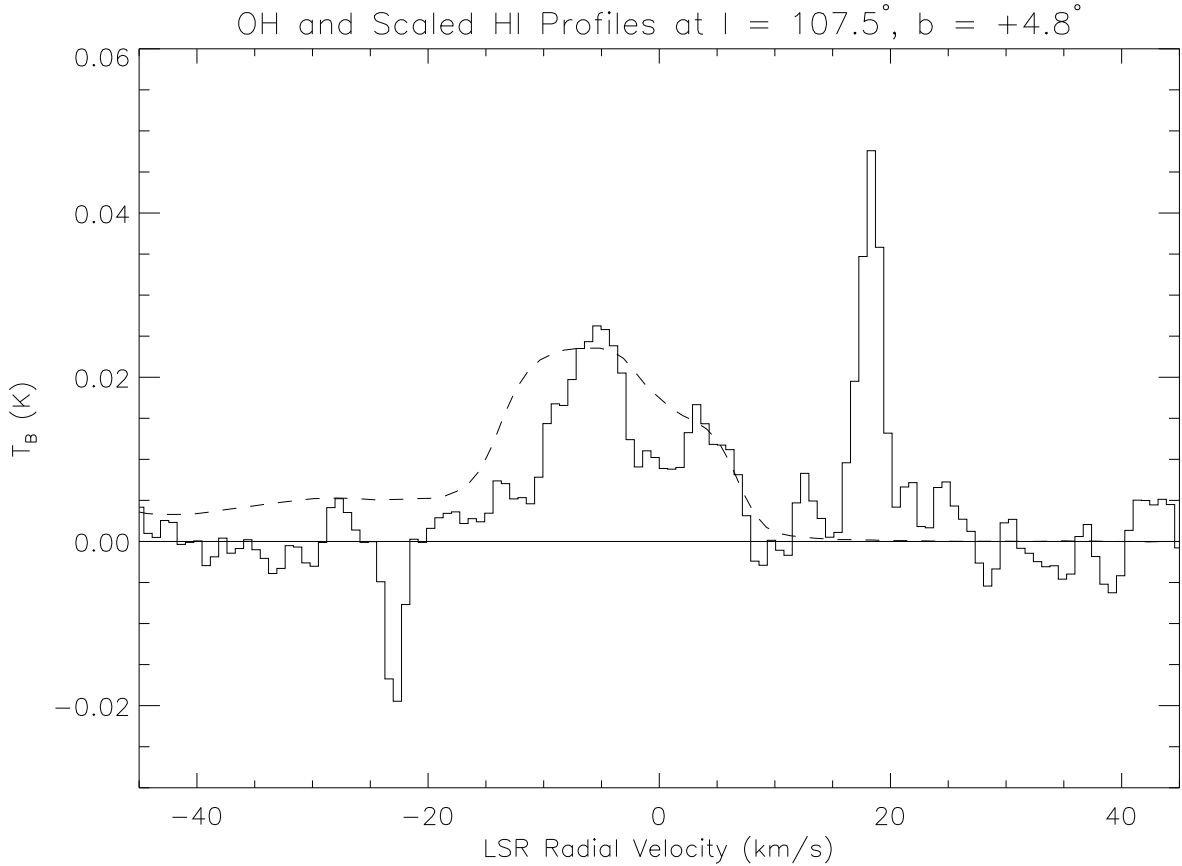}
\caption{
\HI\ profile (dashed) from the LAB survey reduced by a factor of 3500 and plotted along with the OH profile at survey location (0,0) (converted here to units of $T_B$). The narrow spectral features at -23 and +18.5 \kmps\ are spurious, as discussed in Appendix \ref{app:one}.
}
\label{fig:OH-HI}
\end{center}
\end{figure*}

A more detailed comparison of a typical OH velocity profile with the corresponding \HI\ profile structure is made in Figure \ref{fig:OH-HI}, which shows the 21-cm \HI\ profile obtained from the Leiden/Argentine/Bonn (LAB) survey \citep{kal05}\footnote{http://www.astro.uni-bonn.de/webaiub/english /tools\_labsurvey.php} and plotted at a greatly-reduced scale on our most sensitive OH 1667 profile constructed from the average of all observations at $l = 107.5^{\circ},\ b = +4.8^{\circ}$. This OH profile provides an important check on the sensitivity and repeatability of our survey, and it is described in more detail in Appendix \ref{app:one}. The noise level in the OH profile of Figure \ref{fig:OH-HI} is $\approx 3$ mK (in $T_B$ units); the \HI\ profile is essentially noise-free at this scale. Note that the \HI\ profile is much smoother and peaks at just over 80K (in $T_B$).

\begin{figure*}[ht!] 
\begin{center}
\includegraphics[width=\textwidth]{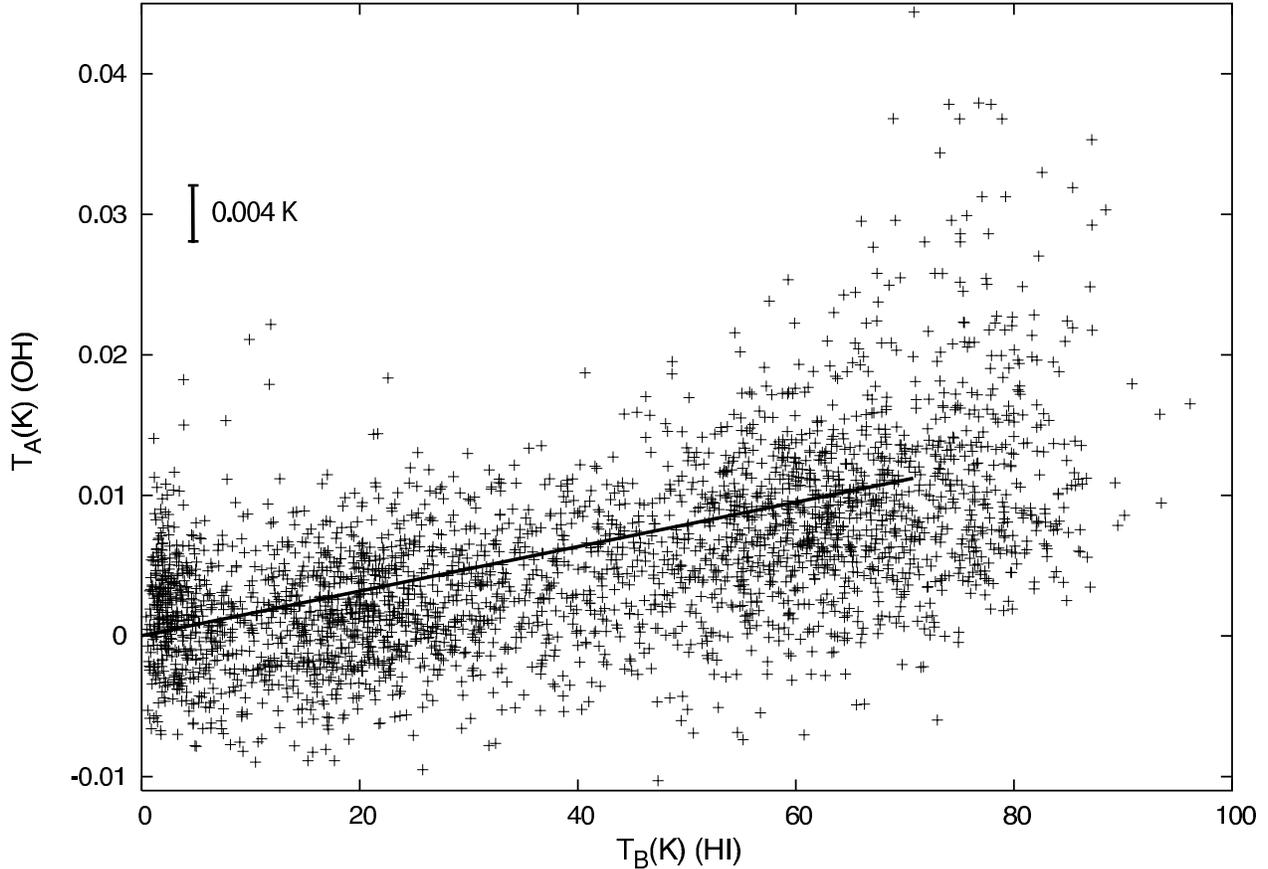}
\caption{
Correlation diagram for corresponding $\approx 1$ \kmps\ velocity channels in the ensemble of spectra from Figures \ref{fig:OH} and \ref{fig:HI}. The velocity range has been limited to that encompassed by the bulk of the signal, $-20 \leq V_{LSR} \leq +15$ \kmps, and in order to avoid the narrow-band interference in the OH spectra at -23 and +18.5 \kmps\ (cf.\ Figure \ref{fig:OH-HI}). The solid line is the relation $T_A$(OH) $ = 1.50 \times 10^{-4}\; T_B$(\HI), or $T_B$(\HI) $ = 6670 \times T_A$(OH). The short vertical bar shows the scale of the rms noise $\Delta T_A^{rms} \approx 3 - 5$ mK in the OH data plotted along the y-axis. The HI data plotted along the x-axis are essentially noise-free at this scale.
}
\label{fig:CorrOH-HI}
\end{center}
\end{figure*}

In order to further investigate the relation between OH and \HI\ profiles, we show in Figure \ref{fig:CorrOH-HI} a correlation diagram for the corresponding $\approx 1$ \kmps\ velocity channels of all pairs of OH and \HI\ spectra in our survey area. With the axis scales used in this figure, the scatter in the plot is dominated by the OH observations ($\Delta T_A^{rms} \approx 3 - 5$ mK); the relative noise in the \HI\ data ($\Delta T_B^{rms} \approx 70$ mK) is negligible in this figure. Figure \ref{fig:CorrOH-HI} shows an approximately-linear relation between the 1667-MHz line of OH and the 21-cm line of \HI\ at corresponding $\approx 1$ \kmps\ channels over the full range of radial velocities, at least up to values of $T_B$(\HI) $\approx 60 - 70$ K. This relation is approximately described by $T_A$(OH) $ \approx 1.50 \times 10^{-4}\; T_B$(\HI), or $T_B$(\HI) $ \approx 6670 \times T_A$(OH) $ \approx 4270 \times T_B$(OH)\footnote{Note that the line shown in Figure \ref{fig:CorrOH-HI} is not the result of a linear least-squares fit; at this point it is merely a convenient parameterization of the data which can be used e.g.\ for planning future observations. More elaborate models will be explored in a future publication.}. Above $T_B$(\HI) $\approx 60 - 70 $ K, the \HI\ line intensity appears to ``saturate'', although the OH signal continues to grow by a further factor of $\approx 3$. At first sight the results suggest that the \HI\ 21-cm line is becoming optically thick above $\approx 60$ K in the solar neighborhood, although \cite{ht03} have shown that the \HI\ gas in the Galaxy shows a wide range of opacities and excitation temperatures. Another possibility is that the atomic gas is disappearing into a molecular phase. In any case, the atomic and molecular components of the gas giving rise to the emission we observe appear to be generally well-mixed in the Galactic ISM on the $\approx 100$ pc scale of our survey. Also, the fact that the OH main-line emission continues to grow after the \HI\ brightness has reached saturation (cf. Figure \ref{fig:CorrOH-HI}), coupled with the greater extent of the OH emission both in space and in velocity compared to CO(1-0) (cf. \S \ref{subsubsec:CO}), suggests that the OH main-line emission may be useful as an alternative indicator for the total column density of \textit{both} the atomic \textit{and} the molecular gas present along the line of sight. Further work on this important topic will be deferred to a later paper.

\subsubsection{Relation of OH to Galactic structure}

The OH profiles are sufficiently extended in space and velocity that they show the effects of Galactic rotation. For distances within about a kiloparsec from the sun, and for uniform circular rotation in a flat plane, the LSR velocity and distance from the sun are related by:
\begin{equation}
\label{eqn:one}
D = - \frac{V_{LSR}}{A \sin(2l)}
\end{equation}
\cite[e.g.][]{bur88}, where $A$ is Oort's constant and $l$ is the Galactic longitude. For $A = 14$ km/s/kpc and $l = 107.5$, the relation is $D = -120 \times V_{LSR}$ (pc/\kmps). However, there is much additional structure in the gas tracers in this region of the Galaxy\footnote{The lower right corner of our survey region extends into the Cepheus complex.}. Also, for such modest radial velocities, we may expect only an approximate relation to exist between velocity and distance. For example, the \HI\ ``shoulder'' in Figure \ref{fig:OH-HI} at around +4 \kmps\ and the associated OH are at an ``anomalous'' positive velocity; their location along the line of sight is uncertain. Proceeding away from the sun, an OH feature appears rather abruptly at -3 \kmps\  (360 pc) in Figure \ref{fig:OH-HI} and is associated with the flat-topped \HI\ profile at around -5 \kmps\ (600 pc). Both the \HI\ and the OH decrease to low values beyond $\approx -10$ \kmps (1.2 kpc).  The ``lump'' in the OH emission in the range $-9 < V_{LSR} < -3$ is a feature at distances of $\approx 350 - 1100$ pc, parts of which likely correspond with the L1204 dark cloud.  For reference, the estimated distance to the H$\alpha$ arc known as Sharpless 140 is about 900 pc \citep{cf74}; this feature is thought to be a PDR on the surface of a molecular cloud excited by one or more nearby B stars. Beyond -20 \kmps\ the OH profile drops below the threshold of detectability; this corresponds to a distance of 2400 pc from the sun, and at $b \approx 5^{\circ}$ the line of sight is now $\approx 200$ pc above the Galactic plane and likely beyond most of the local Galactic gas layer. The \HI\ that persists beyond -20 \kmps\ may belong to the outer Galaxy; the Galactic plane warps upwards to higher latitudes at these longitudes \cite[][\S 7.5]{bur88}. However, any OH emission associated with that \HI\ is apparently too faint to detect, and may in any case have been lost in our baseline subtraction process.

The velocity range where the \HI\ 21-cm line appears to be saturated ($T_B$(\HI)$\; \gtrsim 60$ K) extends from $-10 < V_{LSR} < +5$ \kmps\ over the whole survey area. This gas is widespread but fairly local, within a distance of $\approx 1.2$ kpc from the sun according to equation \ref{eqn:one}. Beyond this distance the linear resolution of our $\approx 30'$ beam exceeds 10 pc and the \HI\ concentrations are likely to be unresolved. The \HI\ features in our survey which are at larger negative velocities (and are therefore likely to be more distant) may still be optically thick, but may not appear so owing to their lower (beam-diluted) profile intensities. Nevertheless, if the OH and \HI\ features in the ISM are generally well-mixed on the $\approx 100$ pc scale as our results suggest, the filling factor will be roughly similar for both lines, thereby preserving the correlation shown in Figure \ref{fig:CorrOH-HI}.

\subsubsection{OH profile amplitude structure; relation to CO}
\label{subsubsec:CO}

\begin{figure*}[ht!] 
\begin{center}
\includegraphics[width=6.0in, angle=-90]{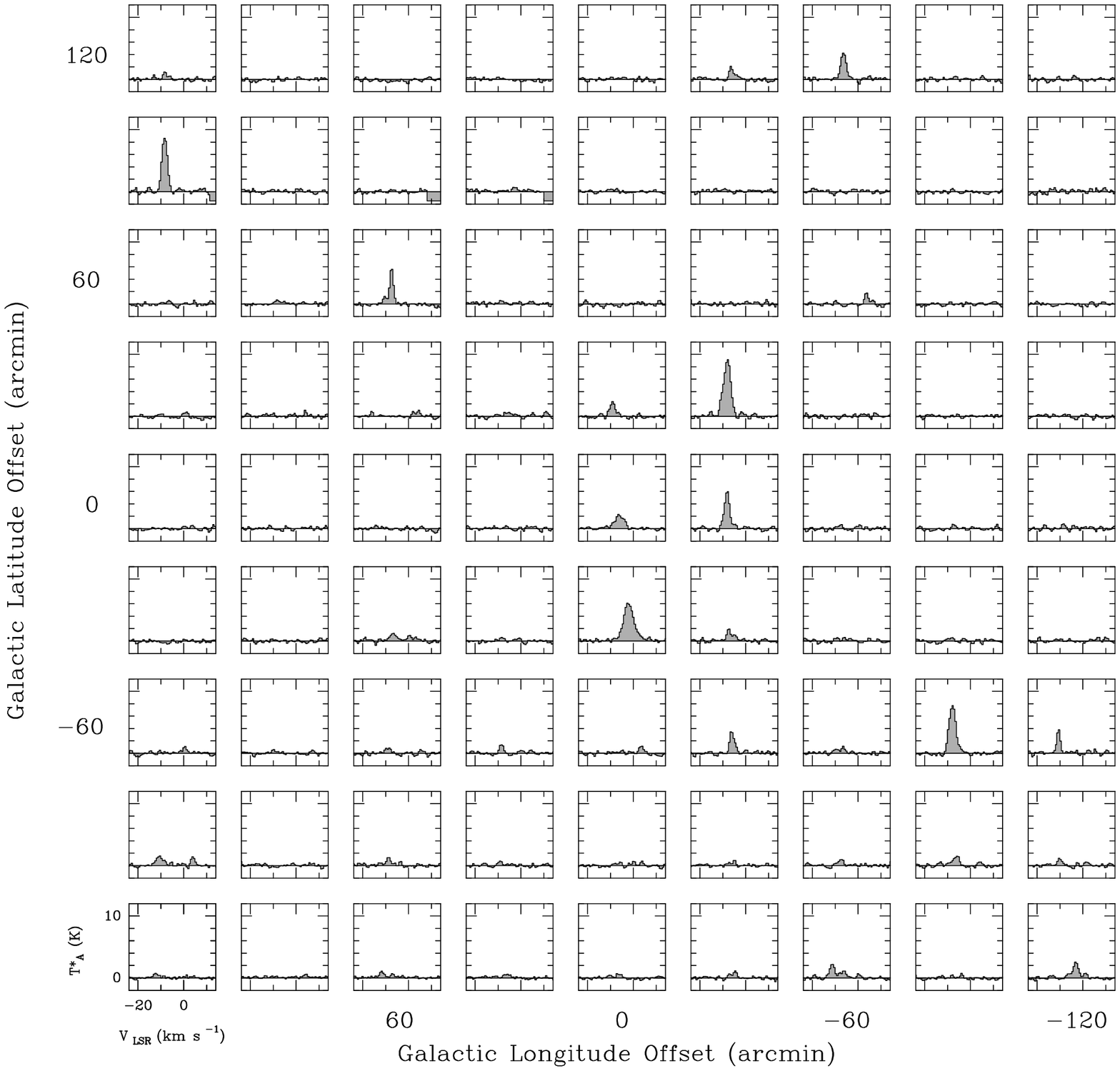}
\caption{
\twCO(1-0) profiles in the survey area. Antenna temperature $T_A^{\ast}$ and LSR velocity scales are given on the panel at the lower left corner.
}
\label{fig:CO}
\end{center}
\end{figure*}

A second feature of the OH spectra is their ``lumpiness'', often showing several identifiable maxima across the full velocity extent. Examples can be noted in Figure \ref{fig:OH} at offset positions $\Delta l, \, \Delta b$ of +120, +90 (3 features); -30, -120 (4 or 5 features); and -30, +30 (2 features). These features are characterized by velocity widths of order 2-3 \kmps\ and velocity separations of order 7-9 \kmps. The peaks of these features are at signal levels of 4-5 $\sigma$ and, as an indication of their physical reality, we note that they occasionally correspond with features seen in CO surveys of this region. In order to illustrate this latter point we have constructed a mosaic of \twCO(1-0) profiles in our survey area using the second-quadrant data cube from the \twCO(1-0) survey by \cite{da01}. This survey was obtained using the 1.2-m CfA millimeter-wave telescope with an angular resolution of $\approx 8.5'$ at 115 GHz. The telescope was initially equipped with a 250 kHz filter bank, but since 1988 this system has also been operated with a 500 kHz filter bank providing a velocity resolution of $\approx 0.65$ \kmps\ and a total velocity coverage of 332 \kmps. The spectra were acquired by position switching. The rms noise level per channel is $\approx 0.31$ K. The CO data cube was convolved in space with a gaussian profile to an angular resolution of $30'$ and resampled to match with our OH observing grid using the tasks SMOOTH and REGRID in the GIPSY reduction package \cite[][]{vtb92}. The spectra were extracted from the CO map and presented in a grid using the GREG package from the GILDAS data analysis system \citep{gf89}. The results (in units of $T_A^{\ast}$) are presented in Figure \ref{fig:CO}.

\begin{figure*}[ht!] 
\begin{center}
\includegraphics[width=\textwidth]{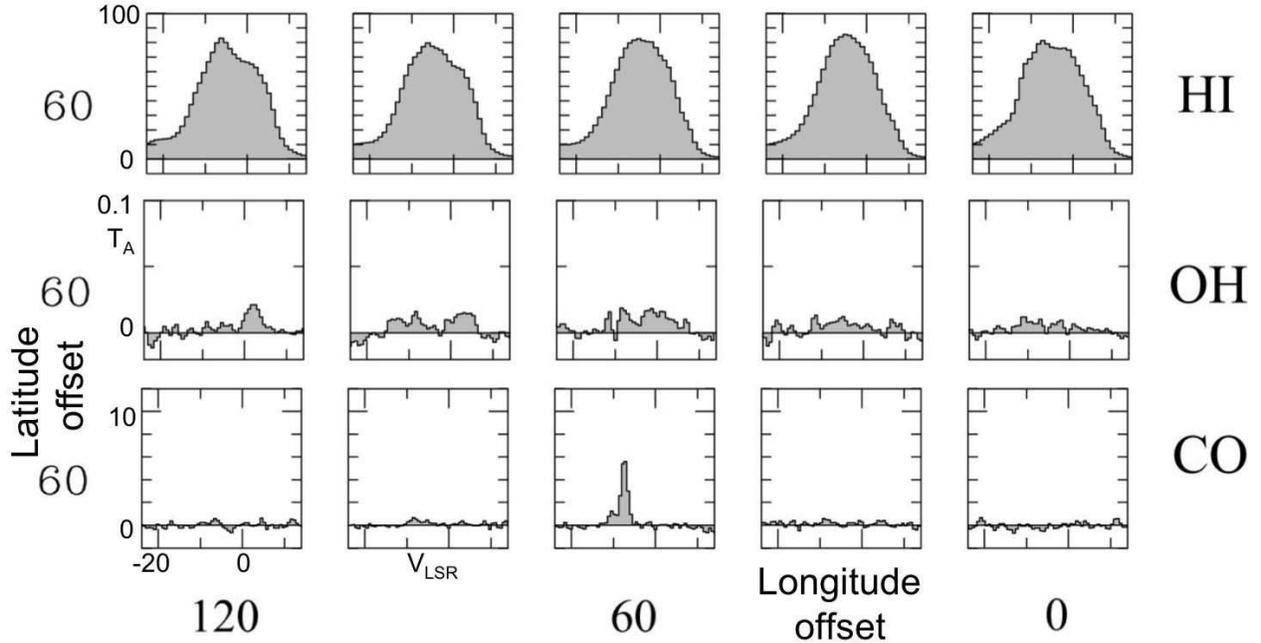}
\caption{
Triptych of profiles from the same location in each mosaic. Amplitude scales from Figures \ref{fig:OH}, \ref{fig:HI}, and \ref{fig:CO} are shown at left. The LSR velocity scale for all spectra is shown on the x-axis of the lower left panel. The temperature scale on the y-axis refers to $T_B$ for \HI, $T_A$ for OH, and $T_A^{\ast}$ for CO. See text for further discussion.
}
\label{fig:triptych} 
\end{center}
\end{figure*}

Many examples of the correspondence between CO and OH emission features can be seen in this figure. However, it is important to note that the converse is not true; the OH profiles show many significant features that have no convincing  counterparts in CO emission at the present levels of sensitivity and resolution. This is clearly illustrated in Figure \ref{fig:triptych} showing sequences of 5 profiles extracted from each mosaic at the same adjacent locations of $0', 30', 60', 90'$, and $+120'$ longitude offset at $+60'$ latitude offset. Note the good correspondence in total velocity extent between the OH (middle row) and the \HI\ profiles (top row). CO features (bottom row) also correspond with OH peaks, but the converse is not true; many OH features appear with few, faint, or no obvious corresponding CO feature. Less than 10\% of the OH features visible in this set of profiles are accompanied by CO emission.

\section{Conclusions}
\label{sec:conclusions}

We have detected ubiquitous 1667 OH line emission in a full-beamwidth single-dish survey of a $\approx 4^{\circ} \times 4^{\circ}$ region located in a direction towards the outer Galaxy at $l \approx 108^{\circ}, b\approx 5^{\circ}$. We have compared the main morphological features of this OH emission to surveys of the 21-cm \HI\ and the 3-mm CO(1-0) lines in the same region. The major results of this comparison are:
\begin{enumerate}
  \item The molecular gas detected in the 18-cm OH line emission is widely distributed both in direction and in velocity, and resembles more closely the distribution of 21-cm \HI\ emission than that of the 3-mm CO line.
  \item For $T_B$(\HI) $\lesssim 60$ K, a linear relation is found between the OH and \HI\ surface brightnesses of the form $T_A$(OH) $ \approx 1.50 \times 10^{-4}\; T_B$(\HI).
  \item Above $T_B$(\HI) $\approx 60$ K, the 21-cm line emission appears to saturate, and few values are recorded above $\approx 90$ K. However, the 1667 MHz OH line brightness continues to grow by a further factor of $\approx 3$.
  \item In contrast to the apparent smoothness of the \HI\ velocity profiles, the OH profiles show a ``lumpy'' structure in velocity, with each feature having a typical amplitude of $T_B \approx 20 - 40$ mK and a velocity FWHM of $\approx 2 - 3$ \kmps.
  \item Less than $10\%$ of the OH features have accompanying CO(1-0) emission,  and the general correlation of OH and CO emission at our $\approx 30'$ survey resolution is poor.
\end{enumerate}

These results raise a number of interesting questions about the physics and chemistry of the ISM, and further analysis in progress. An urgent observational question to be addressed is whether the general spatial correlation of \HI\ and OH emission which we have reported here persists over the entire image of a galaxy. Our current view of the molecular content of galaxies as provided by the 3-mm CO(1-0) emission line is that the surface density of molecular gas is usually highest in the central regions of galaxy disks, and declines rapidly towards the outer parts. On the other hand, the \HI\ distribution often shows a depression in the central regions and declines only slowly with increasing radial distances. What happens to to the main-line OH emission, and why? Observational study of these phenomena will be challenging, but the results promise to be interesting.

\vspace{-0.2in}
\section*{Acknowledgments}

RJA is grateful to Tommy Wiklind for facilitating the initial contact with the Onsala Space Observatory group. We are grateful to the Observatory technical and administrative staff for their capable assistance with our program; Lars Lundahl and Karl-Ake Johansson helped with the observing, and Lars E.B.\ Johansson helped with the initial data reduction. Later stages of the analysis made use of the XS profile analysis software written by P.\ Bergman and installed at STScI by Giuseppe Romeo. We have also used the GILDAS \cite[][]{gf89} and GIPSY \cite[][]{vtb92} software systems. The CO(1-0) data was kindly supplied by Tom Dame.  Background on the early OH surveys of the Galaxy was provided by Miller Goss and Carl Heiles, and Harvey Liszt offered helpful remarks on observations of OH and other molecules in the Galaxy. We are also grateful to the referee for a number of suggestions on an earlier version of this paper.

\appendix


\vspace{-0.2in}
\section{Data consistency and repeatability}
\label{app:one}

\begin{figure*}[ht!]
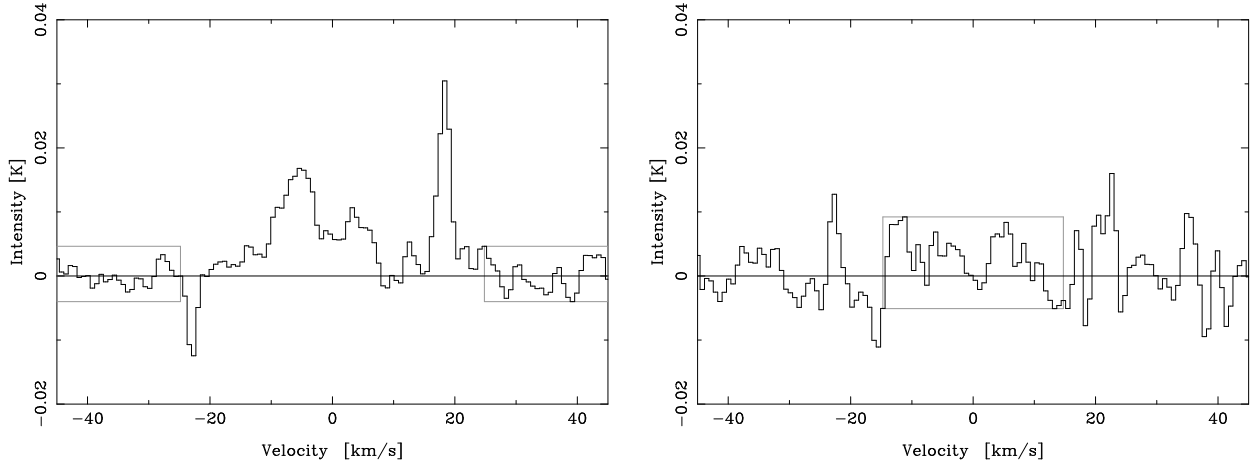
 
\includegraphics[width=2.4in, angle=-90]{fig8a.eps}\hfill\includegraphics[width=2.4in, angle=-90]{fig8b.eps}
\caption{
\textbf{Left:} OSO spectrum of the 1667 OH line in $T_A$ units recorded at $l = 107.5^{\circ},\ b = +4.8^{\circ}$ in 10.0 hours of integration.
\textbf{Right:} The data in the left panel have here been split into two equal time intervals and differenced. \label{fig:refpixel}}
\end{figure*}

At the faint signal levels we are reporting here, there is concern that variable external interference and low-level receiver instability will compromise the integrity of the data. We have evaluated the overall temporal stability of our observations as follows: At regular intervals during both the October and the December 2005 periods, OH spectra were taken at the central (0,0) offset position of the survey grid. These spectra were reduced in the same way with the rest of the data as described in section \ref{sec:observations}. After excising poor spectra, the total usable observing time spent on this central pointing was 10.0 hours, more than 3 times that spent on any other grid point. The average of all reliable data at this offset is shown in the left panel of Figure \ref{fig:refpixel} in $T_A$ units. A straight line was fitted through the spectrum in the velocity ranges $-45 \leq V_{LSR} \leq -25$ \kmps\ and $25 \leq V_{LSR} \leq 45$ \kmps\ (boxes indicated on plot) and removed. The rms noise in the baseline is 2.0 mK, about 1/8 of the peak signal of 17 mK at -5 km/s. The narrow peaks near -23 and +18.5 km/s are due to low, but persistent, levels of interference (probably internal to the observatory, but otherwise of unknown origin); the frequency-switching observing mode employed can result in these peaks appearing to be in ``absorption''. In the right panel of this figure the data have been split into two equal parts of 5 hours each, and differenced. The rms noise in the ``signal'' range $-15 \leq V_{LSR} \leq 15$ km/s (box indicated on plot) is 4.0 mK, exactly twice that in the left panel, as expected.  The narrow peaks near -23 and +18 km/s are less prominent here. We conclude that the features between -20 and +15 km/s in our mosaic of OH spectra in Figure \ref{fig:OH} are repeatable, and that the profile noise is dominated by random statistical fluctuations. 

\end{document}